# Towards Clinical AI Fairness: A Translational Perspective

Mingxuan Liu[1#], Yilin Ning[1#], Salinelat Teixayavong[1], Mayli Mertens[2], Jie Xu[3], Daniel Shu Wei Ting[1,4,5], Lionel Tim-Ee Cheng[6], Jasmine Chiat Ling Ong[7], Zhen Ling Teo[4], Ting Fang Tan[4], Ravi Chandran Narrendar[4], Fei Wang[8], Leo Anthony Celi[9,10,11], Marcus Eng Hock Ong[12,13], Nan Liu[1,5,12,14]*

[1] Centre for Quantitative Medicine, Duke-NUS Medical School, Singapore, Singapore

[2] Department of Philosophy, University of Antwerp, Antwerp, Belgium

[3] Department of Health Outcomes and Biomedical Informatics, University of Florida, Gainesville, FL, USA

[4] Singapore Eye Research Institute, Singapore National Eye Centre, Singapore, Singapore

[5] SingHealth AI Office, Singapore Health Services, Singapore, Singapore

[6] Department of Diagnostic Radiology, Singapore General Hospital, Singapore, Singapore

[7] Department of Pharmacy, Singapore General Hospital, Singapore, Singapore

[8] Department of Population Health Sciences, Weill Cornell Medicine, New York, NY, USA

[9] Institute for Medical Engineering and Science, Massachusetts Institute of Technology, Cambridge, MA, USA

[10] Division of Pulmonary, Critical Care and Sleep Medicine, Beth Israel Deaconess Medical Center, Boston, MA, USA

[11] Department of Biostatistics, Harvard T.H. Chan School of Public Health, Boston, MA, USA

[12] Programme in Health Services and Systems Research, Duke-NUS Medical School, Singapore, Singapore

[13] Department of Emergency Medicine, Singapore General Hospital, Singapore, Singapore

[14] Institute of Data Science, National University of Singapore, Singapore, Singapore

[#] These authors contributed equally

* Corresponding author: Nan Liu, Centre for Quantitative Medicine, Duke-NUS Medical School, 8 College Road, Singapore 169857, Singapore. Phone: +65 6601 6503. Email: liu.nan@duke-nus.edu.sg

## Abstract

Artificial intelligence (AI) has demonstrated the ability to extract insights from data, but the issue of fairness remains a concern in high-stakes fields such as healthcare. Despite extensive discussion and efforts in algorithm development, AI fairness and clinical concerns have not been adequately addressed. In this paper, we discuss the misalignment between technical and clinical perspectives of AI fairness, highlight the barriers to AI fairness' translation to healthcare, advocate multidisciplinary collaboration to bridge the knowledge gap, and provide possible solutions to address the clinical concerns pertaining to AI fairness.

## 1. Introduction

The early days of artificial intelligence (AI) were filled with great aspirations, some of which have now been realized, particularly in the "post-ChatGPT" era.[1-3] In healthcare, data-driven AI models have shown capability in extracting objective evidence from complex and large-scale databases[4, 5]. Yet, algorithms are only as objective as the data that they are based on. Similarly, human judgments are inevitably susceptible to bias in handling sensitive data (e.g., age, gender, race, socio-economic status, weight, sexual orientation) even when these data variables have no objective connection with the outcome of interest[6]. In such a high-stakes field like clinical decision-making, fairness (or absence of bias) is of vital importance. Proper application of AI fairness in clinical algorithmic development could contribute to the reduction of health disparities rather than their escalation[7, 8] but practical implementation is not self-evident.

The practice of medicine has continuously been evolving from "eminence-based" (on the reputation of the physician) to "evidence-based" (on reproducible evidence from well-designed studies). However, in the absence of sufficient resources, the evidence is often gathered from a possibly skewed representation of the underlying population, e.g., in terms of race or age groups. The emerging data-driven practice in medical decision-making based on big data may partially mitigate the risk of bias, but if not carefully designed, such an approach can still lead to unfair decisions[9]. For example, the online Kidney Donor Profile Index (KDPI) calculator used by the US Organ Procurement & Transplantation Network predicts higher risks of kidney graft failure for black donors than for non-black donors when all other conditions are identical, resulting in fewer eligible organ sources from black donors. This consequently leads to a lower chance of the match for black recipients[10], since blood type and Human Leukocyte Antigens (HLA) matching are more likely in the same race group.

Such risk of bias is not automatically mitigated by using more complex AI algorithms or a large amount of data[11]. In one example, questionable differences are observed in AI-based survival prediction after liver transplantation among gender and race groups[12, 13], which can bias clinical decisions and allocation of scarce healthcare resources against certain patient subgroup(s) simply because of the traits they were born with. Such models that hold discrimination against demographic subgroups are far away from the justice required in delivering equal well-being in healthcare. We therefore agree it is essential to develop fair models for data-driven clinical decision-making but current AI fairness research may not be well-adaptable for clinical settings.

In recent years, with growing public awareness of bias in AI models in real-life tasks such as face recognition[14] and prediction of recidivism[15], AI researchers have developed a large number of qualitative and quantitative approaches to evaluate and ensure fairness in model development[16, 17]. However, due to the knowledge gap amongst AI researchers and clinicians, AI fairness studies tend to either focus on abstractive conceptualization or technical modeling regarding fairness metrics. While these aspects are highly important, it is unclear how they can be applied to healthcare. As illustrated in Figure 1, in this paper we discuss the misalignment of current AI fairness research with practical clinical concerns and highlight the obstacles to AI fairness adaptation.

## 2. AI fairness from a technical perspective

Fair AI has been associated with a variety of technical properties and capabilities. It is believed that AI is capable of making accurate predictions. Additionally, AI is expected to remain robust against the cognitive bias and prejudice that humans experience when making judgments, and is even capable of detecting bias that humans cannot recognize[16, 17].

### 2.1 Bias and fairness

Bias and fairness are two concepts that usually oppose each other: a decision is not fair if it is biased towards (or against) any individual or subpopulation. In the context of data and modeling, bias refers to the discrimination against certain groups or individuals in terms of sensitive variables (e.g., age, gender, socio-economic status, etc.), and is contrary to fairness[18]. Although intuitively correct, such a definition of fairness is difficult to incorporate into AI model development as it is difficult to quantify and, more importantly, as it may conflict with clinical goals. In the development pipeline of an AI model, which typically involves data collection, model training, evaluation, and validation, bias (and therefore unfairness) can occur at any stage for a variety of reasons, sometimes in an imperceptible manner.[19]

First, any historical (or current) bias in medical practice will be reflected in medical records, e.g., the bias in prescribing pain treatments for patients from different race groups[20], which will bias the resulting prediction models in similar ways unless carefully handled during model development. Data under-representation[21] is another common source of data bias that arises from inappropriate data collection and sampling, where certain subgroups constitute a smaller proportion of the sample than they are prevalent in the underlying population. This induces data imbalance that can lead to bias in model-based decision-making against groups with limited information. Such bias occurred in the landmark Framingham Heart Study, which failed to fully represent subpopulations with different disease etiology and risk factors and consequently had poor predictive performance in some underrepresented groups (e.g, African Americans)[22]. Data bias may be amplified by inappropriate data pre-processing, including but not limited to exclusion of incomplete records when missing was not at random, or a naive combination of datasets from different sources without accounting for overlapping subjects. All possible sources of data bias should be proactively identified and addressed during the early stages of AI model development before it becomes an impediment to algorithmic fairness.

In addition to data bias, inappropriate model development steps (e.g., unjustifiable use of sensitive variables such as gender and race in decision-making) can amplify existing bias or introduce new bias in the resulting AI models, resulting in algorithm bias that is another prevalent source of AI unfairness[8]. The use of black-box AI models, especially complex deep learning models, exacerbates algorithm bias by making it more difficult to detect. Algorithm bias can be mitigated, but often at the expense of model performance, e.g., when intentionally excluding sensitive variables that can add information for outcome prediction in the development data. This makes it difficult to develop and implement completely fair AI in pragmatic healthcare practice.

## 2.2 Fairness metrics in AI literature

Many quantitative metrics have been developed to measure fairness in AI, mostly from the perspective of "equality" [16, 17, 23]: a fair model should have "equal performance" in subgroups with respect to sensitive variables. Generally, there are two types of metrics with distinct mechanisms, i.e., group-based and individual-based[16, 17, 23]. Group-based metrics measure the consistency of model performance (e.g., using the confusion matrix or calibration) across subgroups defined by sensitive variables, and a fair model is expected to behave similarly among subgroups. Some widely used individual-based metrics are fairness through awareness[24] which assumes that observations with similar conditions should have similar predictions, and counterfactual fairness[25] expecting the predicted outcome for an individual to remain unchanged after changing some sensitive variable.

There is currently no consensus on choosing between group-based and individual-based metrics as they quantify different aspects of fairness[26-28], and it is impossible to simultaneously improve model fairness with respect to both types of metrics due to their mathematical definitions[26, 27]. Moreover, the ethical assumptions underlying these metrics require further clarifications on their practical implications, and failing to account for either individual- or group-based fairness seems unethical[27]. The lack of unified fairness metrics and clear application guidelines cast confusion on both AI researchers and clinical decision-makers.

## 2.3 Methods to detect, prevent, and mitigate bias

To ensure the fairness of AI models, each step of the modeling pipeline should be self-motivated and aware of fairness[23], even for data exploration[29]. A detailed description of datasets (e.g., time-period and site information for data collection) can provide evidence to detect potential bias, e.g., under-representation of any subpopulation[21], which enables early bias prevention. A simple way to resolve such data bias is to collect or request additional data, but this is not always feasible due to regulations and legislation. In this case, AI researchers can pre-process existing data using appropriate sampling methods, such as preferential sampling[30], to better represent the underlying population, and use regular methods to develop models from the adjusted dataset.

In addition to the pre-processing approach described above, there is a rich body of research on methods to mitigate data bias in-process during model development, using the fairness metrics described in the previous subsection as bias-monitoring and fairness-evaluation tools. Conditioning on the sensitive variables yields two general types of methods based on subgroup analysis: either building separate models for each subgroup with a customization thinking, or constructing a unified model and respectively evaluating it with subgroup data. Fairness metrics can also serve as constraints for these models, in addition to loss functions that measure the accuracy of predictions used in conventional model development. Another approach is representation learning, which ensures independency between predictions and sensitive variables by generating latent representation with information on sensitive variables eliminated[31]. Another common method of representation learning is adversarial thinking, which consists of instructing generative adversarial network (GAN) models to generate latent representations based on which even the models cannot detect sensitive information[32].

## 3. AI fairness from a clinical perspective

Existing AI fairness methodological research and applications are often in areas where "fairness" is simply defined as equality among subpopulations[16, 17], e.g., the equal predicted probability of recidivism among race groups. In clinical applications, however, it may be plausible to assume a higher likelihood of certain diagnoses or medical/social complications for specific subgroups of patients: elder people with a higher risk for Alzheimer's disease, people with obesity at higher risk for metabolic disease, and a higher likelihood of long-term health conditions in lower socio-economic groups, etc. Hence, there is a prominent gap in the understanding of "fairness" between AI developers and healthcare providers, which may explain the limited clinical applications of AI fairness. In this section, we point out the potential hurdles and challenges in AI fairness in healthcare to promote its future applications in clinical settings.

### 3.1 Hurdles for evaluating fairness in healthcare

On top of these AI fairness metrics aforementioned in section 2.2, the uncertainty lies where the mechanisms behind the fairness metrics can be plausible from the perspective of healthcare. For example, counterfactual fairness assumes that the prediction should not change for an individual after changing the value of some sensitive variable, e.g., when a female changes to a male with all other variables unchanged. This may be plausible when predicting the likelihood of being hired by a company, but less so in clinical contexts with natural biological differences between female and male[33]. Therefore, artificially changing one sensitive variable while leaving others unchanged would construct a "phantom" and such a comparison between an individual and a corresponding "phantom" seems unreliable.

Moreover, without a unified formal definition of "fairness" in clinical applications, every individual may have different perceptions of fairness that varies by context: for patients, individual-based fairness may be more relevant, whereas for the management team of a hospital group-based fairness may be preferable. Additionally, for each type of fairness definition (i.e., group-based and individual-based), the presence of multiple metrics burdens the application of AI fairness in healthcare, as there is no clear consensus on the optimal way to quantify fairness, and different approaches can produce inconsistent results[23]. Though there have been several review papers discussing the relationships and differences between these metrics, they do not provide practical guidelines to address specific clinical needs[34]. One possible solution is to address all metrics together via methods like multi-task learning[35]. However, due to the trade-offs between the metrics, it is mathematically impossible to optimize all metrics in the same model, except in highly restrictive cases[23].

In addition, group-based metrics are "secondary" metrics derived from primary performance metrics (e.g., true positive rates and false positive rates), therefore aiming for equality in such metrics among sensitive groups can have an unclear impact on overall model performance. Differences in metrics among sensitive groups are often based on single-point estimates, which lack objective cut-offs and can lead to overclaims of unfairness. Several hypothesis testing methods[36-40] have been proposed to statistically assess the presence of a difference, e.g., by using the permutation test[36]. However, since statistical significance is based on p-values that are affected by sample sizes, a statistically significant difference may not be large enough in magnitude to be clinically significant.[41] It would be relevant to incorporate such considerations when modifying existing fairness metrics or devising new ones for clinical AI models.

## 3.2 Differences or bias?

In the clinical context, differences and biases can be difficult to disentangle, and failure to distinguish them could lead to negative consequences[42]. On one hand, claims of differences can be biased if they lack solid justification; for instance, genetic differences between races in relation to certain diseases can be controversial[9]. In addition, when the biomedical differences have been identified, e.g., males hold a higher risk of non-small-cell lung cancer than females[43, 44], distinguishing bias from differences remains another challenge, as it is reasonable for a model to perform differently among subgroups with significant differences.

On the other hand, simple claims of biased predictions could conceal the real problems that merit further investigations. After reporting bias in their models, most studies either stopped there or tried to mitigate the bias directly, such as by re-training models to yield similar performance for different subgroups. For some studies, this only resulted in their models being fragile and having limited generalizability[45]. However, in clinical settings, identifying the underlying causes makes the claimant solid. In the study of breast cancer, in the face of disparities in model performance, tremendous efforts have been taken to find the underlying reasons. Compared with Singaporean females, Malaysian females tended to conceal their diseases and resisted medical examinations, causing them to be admitted to hospitals at a more severe stage[46].

## 3.3 The problematic assumption underlying current fair AI methodologies

Superficially considering difference as bias, current methodologies of AI fairness can mainly contribute to solving clinical questions that particularly assume "equality" as evidence of non-bias (fairness), such as equal chance of receiving treatment among pre-defined groups (e.g., by age or gender). However, this hypothesis makes one important assumption that may be problematic in real practice. The assumption that treatment is equally suitable for all groups, warranting equal chances for individuals with similar characteristics except for their group memberships, could be controversial and/or even clinically wrong. It is controversial for "race" since it is still debatable if there are no disease-related genetic differences in race groups, and irrational for other variables like "age". Pursuing equality in treatment regardless of age group neglects important dimensions of medical practice including dignity preservation and quality of life optimization. Thus, definitions of AI fairness must be contextualized to both clinical scenarios and society, which would inevitably involve different sets of assumptions, and be informed by real-life feasibility.

## 3.4 Rethinking "sensitive variables" with respect to healthcare scenarios

Variables such as age, gender, race, social status, marital status, and disability status are considered sensitive in general fair AI studies[16, 47], and a fair decision-making process is expected to be free from the influence of such information. However, as discussed above, some of these variables are highly relevant to disease diagnosis, treatment decisions and prognosis, and therefore cannot be hidden from clinical decision-making and healthcare resource allocation. Furthermore, as AI fairness is often improved at the cost of predictive performance (see Section 2.1), diagnostic tools versus treatment decision support tools (evidence-based medicine) may subscribe to different levels of "clinical fairness" depending on the level of accuracy required.

When developing fair AI for clinical outcomes, the set of relevant sensitive variables should be carefully re-evaluated for each application. Race is particularly challenging to handle, as it can be associated with systemic bias that still affects clinical practice, or genuine biological and/or sociological differences among subpopulations. The aforementioned racial bias in pain treatment is a typical example of the former, whereas the later can be controversial and requires additional investigations. For example, while the KDPI score to predict kidney graft failure was criticized for predicting a higher risk for black donors, further investigations revealed that this may be justified by genetic differences common between black and non-black[48]. However, instead of using race as an easy surrogate, it is preferable to replace the proxy with the underlying factors (in this example the genetic factors) affecting the outcome[22]. Such follow-up studies can also help identify modifiable factors to improve healthcare outcomes, instead of passively associating inferior outcomes with some racial groups. Therefore, AI fairness in healthcare must account for clinical justifications rather than simply applying pre-processing or in-processing exclusion for pre-defined sensitive variables. Moreover, failure to include sensitive variables in a model when they are understood to be clinically relevant can lead to bias and distrust.

When tailoring AI fairness to clinical outcomes, in addition to minimizing bias among subgroups defined by sensitive variables, a fair treatment decision requires other crucial concerns that are not applicable to general AI fairness research. These concerns include over-treatment (over-diagnosis) or under-treatment (under-diagnosis[49]), patients' implicit considerations of interests such as end-of-life care preferences, clinicians'/patients' prejudice towards a specific treatment, and lack of AI digital literacy that may limit lower-resource communities from adopting and benefitting from AI, etc. These questions are currently overshadowed by sensitive variables, hindering the applications of well-established methodologies in fair AI, due to the double focus.

## 4. Clinically meaningful AI fairness

With the intertwinement of data and algorithmic bias, and with the extensive works that have been done in the AI community, many things can make AI fairness more meaningful in clinical settings. One stands at the data side, that is, before addressing the subsequent algorithmic bias, the information stored in the data should be unbiased and fully understood; on the other side, the pursuit of clinical fairness should not be isolated from clinicians who can provide direct evaluation along the way.

### 4.1 Addressing data bias before algorithmic bias

Before developing AI models, detailed descriptive analysis should be conducted to fully understand and address any bias present in the current data, or at least make explicit the implication of data bias on model interpretation when the bias cannot be resolved due to practical constraints. Data under-representation is a common type of data bias, and one way to resolve it is data sharing, which helps better represent subgroups of interest by combining data from multiple sources[6, 50]. When data cannot be directly shared due to the vital importance of data privacy, information may be indirectly combined across relevant sites via federated learning to increase data volume and the number of predictors. However, extra caution is needed to ensure fairness in federated learning without bias towards any local sites.

After detecting and mitigating bias in the data, researchers also need to reframe the concept of "fairness" in the specific context of the clinical question. Focussing on principles of equity rather than equality may push concepts of fairness beyond the commonly discussed sensitive variables in general fair AI research. This will likely require inclusion of patients' preferences,Ffor example, patients' prior preference towards treatment when analyzing the possibility of over-treatment. In addition, when feasible, developers ought to avoid using unmodifiable factors (e.g., race) as surrogates for biological or socio-economical factors (e.g., genetic information) that can better explain the causal mechanism for the outcome of interest[51].

## 4.2 Clinicians in the loop with explainable AI

Clinical AI fairness is an interdisciplinary research topic that requires inputs from AI researchers, clinicians, and preferably ethicists. Some of the concerns regarding clinical AI fairness have already been discussed and addressed in medical ethics and bioethics literature, calling for effective cross-disciplinary communication. During the process of mutual understanding between human and AI models, with human experts in the loop, i.e., clinicians, we can not only proactively identify the bias, but also appropriately differentiate bias from "clinically meaningful" differences, thus setting the right objective for AI models. In addition, with clinicians evaluating the algorithms that claimed to handle bias, standard clinical significance regarding fairness can be put forward under the clinical common sense. Thereafter, only the models with clinically significant bias should be adjusted, avoiding over-adjustment or over-claim of bias.

To align the objectives of AI developers and clinicians, it is necessary to establish a flow of information between the two parties. Explainable AI can contribute to such communication since it provides clinicians with the capability of interpreting models[52, 53] and the possibility of giving feedback to the AI developers. In a simplified and practical manner, AI models could display different outputs for clinicians, toggled according to various inputs being considered. As an example, clinical infrastructure may display recommended diagnosis/treatment based on algorithms developed from all variables (including sensitive ones) or selected variables (excluding specific sensitive variables such as age and race). By comparing the two, the physician in the loop can quickly determine whether the sensitive variables have affected the AI algorithm recommendation, and how. Having a better understanding of the model's decision-making process could enable clinicians to improve the model's accuracy, and clinicians would also guide the algorithms in a more equitable direction.

## 5. Conclusions

Current AI fairness research may not be readily adaptable to clinical settings. With the discussion of misalignment between the technical and clinical perspectives, we highlighted the obstacles to clinical AI fairness translation, and made a case for multidisciplinary research involving clinicians, AI researchers, social scientists, and philosophers.

**References**

1. Turing AM. Computing machinery and intelligence. *Parsing the turing test*. Springer; 2009:23-65.
2. LeCun Y, Bengio Y, Hinton G. Deep learning. *Nature*. 2015;521(7553):436-444.
3. Haenlein M, Kaplan A. A brief history of artificial intelligence: On the past, present, and future of artificial intelligence. *California management review*. 2019;61(4):5-14.
4. Yang YC, Islam SU, Noor A, Khan S, Afsar W, Nazir S. Influential Usage of Big Data and Artificial Intelligence in Healthcare. *Comput Math Methods Med*. 2021;2021:5812499. doi:10.1155/2021/5812499
5. Bohr A, Memarzadeh K. The rise of artificial intelligence in healthcare applications. *Artificial Intelligence in Healthcare*. 2020:25-60.
6. Mertens M. *Bias in Medicine. The Rowman & Littlefield Handbook of Bioethics*. Rowman & Littlefield; 2022.
7. Fletcher RR, Nakeshimana A, Olubeko O. Addressing Fairness, Bias, and Appropriate Use of Artificial Intelligence and Machine Learning in Global Health. *Front Artif Intell*. 2020;3:561802. doi:10.3389/frai.2020.561802
8. Tsai TC, Arik S, Jacobson BH, et al. Algorithmic fairness in pandemic forecasting: lessons from COVID-19. *npj Digital Medicine*. 2022/05/10 2022;5(1):59. doi:10.1038/s41746-022-00602-z
9. Vyas DA, Eisenstein LG, Jones DS. Hidden in Plain Sight — Reconsidering the Use of Race Correction in Clinical Algorithms. *New England Journal of Medicine*. 2020;383(9):874-882. doi:10.1056/NEJMms2004740
10. Doshi MD, Schaubel DE, Xu Y, Rao PS, Sung RS. Clinical Utility in Adopting Race-free Kidney Donor Risk Index. *Transplant Direct*. Jul 2022;8(7):e1343. doi:10.1097/txd.0000000000001343
11. Volovici V, Syn NL, Ercole A, Zhao JJ, Liu N. Steps to avoid overuse and misuse of machine learning in clinical research. *Nature Medicine*. 2022/10/01 2022;28(10):1996-1999. doi:10.1038/s41591-022-01961-6
12. Lai JC, Pomfret EA, Verna EC. Implicit bias and the gender inequity in liver transplantation. *Am J Transplant*. Jun 2022;22(6):1515-1518. doi:10.1111/ajt.16986
13. Wingfield LR, Ceresa C, Thorogood S, Fleuriot J, Knight S. Using Artificial Intelligence for Predicting Survival of Individual Grafts in Liver Transplantation: A Systematic Review. *Liver Transpl*. Jul 2020;26(7):922-934. doi:10.1002/lt.25772
14. Menezes HF, Ferreira ASC, Pereira ET, Gomes HM. Bias and Fairness in Face Detection. 2021:247-254.
15. Angwin J, Larson J, Mattu S, Kirchner L. Machine bias: There's software used across the country to predict future criminals. and it's biased against blacks. https://www.propublica.org/article/machine-bias-risk-assessments-in-criminal-sentencing
16. Caton S, Haas C. Fairness in Machine Learning: A Survey. 2020:arXiv:2010.04053. Accessed October 01, 2020. https://ui.adsabs.harvard.edu/abs/2020arXiv201004053C
17. Mehrabi N, Morstatter F, Saxena N, Lerman K, Galstyan A. A Survey on Bias and Fairness in Machine Learning. *ACM Comput Surv*. 2021;54(6)doi:10.1145/3457607
18. Mitchell S, Potash E, Barocas S, D'Amour A, Lum K. Algorithmic Fairness: Choices, Assumptions, and Definitions. *Annual Review of Statistics and Its Application*. 2021;8(1):141-163. doi:10.1146/annurev-statistics-042720-125902
19. DeCamp M, Lindvall C. Latent bias and the implementation of artificial intelligence in medicine. *J Am Med Inform Assoc*. Dec 9 2020;27(12):2020-2023. doi:10.1093/jamia/ocaa094
20. Todd KH, Deaton C, D'Adamo AP, Goe L. Ethnicity and analgesic practice. *Annals of Emergency Medicine*. 2000/01/01/ 2000;35(1):11-16. doi:https://doi.org/10.1016/S0196-0644(00)70099-0

21.	de Hond AAH, Leeuwenberg AM, Hooft L, et al. Guidelines and quality criteria for artificial intelligence-based prediction models in healthcare: a scoping review. *npj Digital Medicine*. 2022/01/10 2022;5(1):2. doi:10.1038/s41746-021-00549-7
22.	Genovese G, Friedman DJ, Ross MD, et al. Association of trypanolytic ApoL1 variants with kidney disease in African Americans. *Science*. Aug 13 2010;329(5993):841-5. doi:10.1126/science.1193032
23.	Xu J, Xiao Y, Wang WH, et al. Algorithmic fairness in computational medicine. *eBioMedicine*. 2022;84doi:10.1016/j.ebiom.2022.104250
24.	Dwork C, Hardt M, Pitassi T, Reingold O, Zemel R. Fairness through Awareness. presented at: Proceedings of the 3rd Innovations in Theoretical Computer Science Conference; 2012; https://doi.org/10.1145/2090236.2090255 {New York, NY, USA}
25.	Kusner M, Loftus J, Russell C, Silva R. Counterfactual fairness. 2017;
26.	Kleinberg J, Mullainathan S, Raghavan M. Inherent Trade-Offs in the Fair Determination of Risk Scores. 2016:arXiv:1609.05807. Accessed September 01, 2016. https://ui.adsabs.harvard.edu/abs/2016arXiv160905807K
27.	Binns R. On the Apparent Conflict between Individual and Group Fairness. *Fat\* '20*. 2020:514–524. doi:10.1145/3351095.3372864
28.	Lee MSA, Floridi L, Singh J. Formalising trade-offs beyond algorithmic fairness: lessons from ethical philosophy and welfare economics. *AI and Ethics*. 2021/11/01 2021;1(4):529-544. doi:10.1007/s43681-021-00067-y
29.	Russo D, Zou J. How much does your data exploration overfit? Controlling bias via information usage. 2015:arXiv:1511.05219. Accessed November 01, 2015. https://ui.adsabs.harvard.edu/abs/2015arXiv151105219R
30.	Cecconi L, Grisotto L, Catelan D, Lagazio C, Berrocal V, Biggeri A. Preferential sampling and Bayesian geostatistics: Statistical modeling and examples. *Statistical Methods in Medical Research*. 2016;25(4):1224-1243. doi:10.1177/0962280216660409
31.	Moyer D, Gao S, Brekelmans R, Steeg GV, Galstyan A. Invariant Representations without Adversarial Training. *Nips'18*. 2018:9102–9111.
32.	Li J, Ren Y, Deng K. FairGAN: GANs-Based Fairness-Aware Learning for Recommendations with Implicit Feedback. *Www '22*. 2022:297–307. doi:10.1145/3485447.3511958
33.	Butler AA, Menant JC, Tiedemann AC, Lord SR. Age and gender differences in seven tests of functional mobility. *Journal of NeuroEngineering and Rehabilitation*. 2009/07/30 2009;6(1):31. doi:10.1186/1743-0003-6-31
34.	Mbakwe AB, Lourentzou I, Celi LA, Wu JT. Fairness metrics for health AI: we have a long way to go. *EBioMedicine*. Mar 14 2023;90:104525. doi:10.1016/j.ebiom.2023.104525
35.	Caruana R. Multitask Learning. *Machine Learning*. 1997/07/01 1997;28(1):41-75. doi:10.1023/A:1007379606734
36.	DiCiccio C, Vasudevan S, Basu K, Kenthapadi K, Agarwal D. Evaluating Fairness Using Permutation Tests. *Kdd '20*. 2020:1467–1477. doi:10.1145/3394486.3403199
37.	Taskesen B, Blanchet J, Kuhn D, Nguyen VA. A Statistical Test for Probabilistic Fairness. *FAccT '21*. 2021:648–665. doi:10.1145/3442188.3445927
38.	DiCiccio C, Vasudevan S, Basu K, Kenthapadi K, Agarwal D. Evaluating Fairness Using Permutation Tests. presented at: Proceedings of the 26th ACM SIGKDD International Conference on Knowledge Discovery & Data Mining; 2020; Virtual Event, CA, USA. https://doi.org/10.1145/3394486.3403199
39.	Taskesen B, Blanchet J, Kuhn D, Nguyen VA. A Statistical Test for Probabilistic Fairness. presented at: Proceedings of the 2021 ACM Conference on Fairness, Accountability, and Transparency; 2021; Virtual Event, Canada. https://doi.org/10.1145/3442188.3445927
40.	Gursoy F, Kakadiaris IA. Error Parity Fairness: Testing for Group Fairness in Regression Tasks. 2022:arXiv:2208.08279. Accessed August 01, 2022. https://ui.adsabs.harvard.edu/abs/2022arXiv220808279G
41.	Kazdin AE. The meanings and measurement of clinical significance. *J Consult Clin Psychol*. Jun 1999;67(3):332-9. doi:10.1037//0022-006x.67.3.332

42. de Kanter A-FJ, van Daal M, de Graeff N, Jongsma KR. Preventing Bias in Medical Devices: Identifying Morally Significant Differences. *The American Journal of Bioethics*. 2023/04/03 2023;23(4):35-37. doi:10.1080/15265161.2023.2186516
43. Ragavan M, Patel MI. The evolving landscape of sex-based differences in lung cancer: a distinct disease in women. *European Respiratory Review*. 2022;31(163):210100. doi:10.1183/16000617.0100-2021
44. Visbal AL, Williams BA, Nichols III FC, et al. Gender differences in non–small-cell lung cancer survival: an analysis of 4,618 patients diagnosed between 1997 and 2002. *The Annals of thoracic surgery*. 2004;78(1):209-215.
45. Cotter A, Gupta M, Jiang H, et al. Training Well-Generalizing Classifiers for Fairness Metrics and Other Data-Dependent Constraints. presented at: Proceedings of the 36th International Conference on Machine Learning; 2019; Proceedings of Machine Learning Research. https://proceedings.mlr.press/v97/cotter19b.html
46. Ng CWQ, Lim JNW, Liu J, Hartman M. Presentation of breast cancer, help seeking behaviour and experience of patients in their cancer journey in Singapore: a qualitative study. *BMC Cancer*. 2020/11/10 2020;20(1):1080. doi:10.1186/s12885-020-07585-8
47. Mehrabi N, Morstatter F, Saxena N, Lerman K, Galstyan A. A Survey on Bias and Fairness in Machine Learning. *ACM Comput Surv*. 2021;54(6):Article 115. doi:10.1145/3457607
48. Freedman BI, Pastan SO, Israni AK, et al. APOL1 Genotype and Kidney Transplantation Outcomes From Deceased African American Donors. *Transplantation*. Jan 2016;100(1):194-202. doi:10.1097/tp.0000000000000969
49. Seyyed-Kalantari L, Zhang H, McDermott MBA, Chen IY, Ghassemi M. Underdiagnosis bias of artificial intelligence algorithms applied to chest radiographs in under-served patient populations. *Nature Medicine*. 2021/12/01 2021;27(12):2176-2182. doi:10.1038/s41591-021-01595-0
50. Seastedt KP, Schwab P, O'Brien Z, et al. Global healthcare fairness: We should be sharing more, not less, data. *PLOS Digital Health*. 2022;1(10):e0000102. doi:10.1371/journal.pdig.0000102
51. Brems JH, Ferryman K, McCormack MC, Sugarman J. Ethical Considerations Regarding the Use of Race in Pulmonary Function Testing. *CHEST*. 2022;162(4):878-881. doi:10.1016/j.chest.2022.05.006
52. Murdoch WJ, Singh C, Kumbier K, Abbasi-Asl R, Yu B. Definitions, methods, and applications in interpretable machine learning. *Proc Natl Acad Sci USA*. 2019/10/29/ 2019;116(44):22071-22080. doi:10.1073/pnas.1900654116
53. Lundberg SM, Lee S-I. A unified approach to interpreting model predictions. presented at: Proceedings of the 31st International Conference on Neural Information Processing Systems; 2017; Long Beach, California, USA.

Figure 1: Conceptual model towards clinical AI fairness

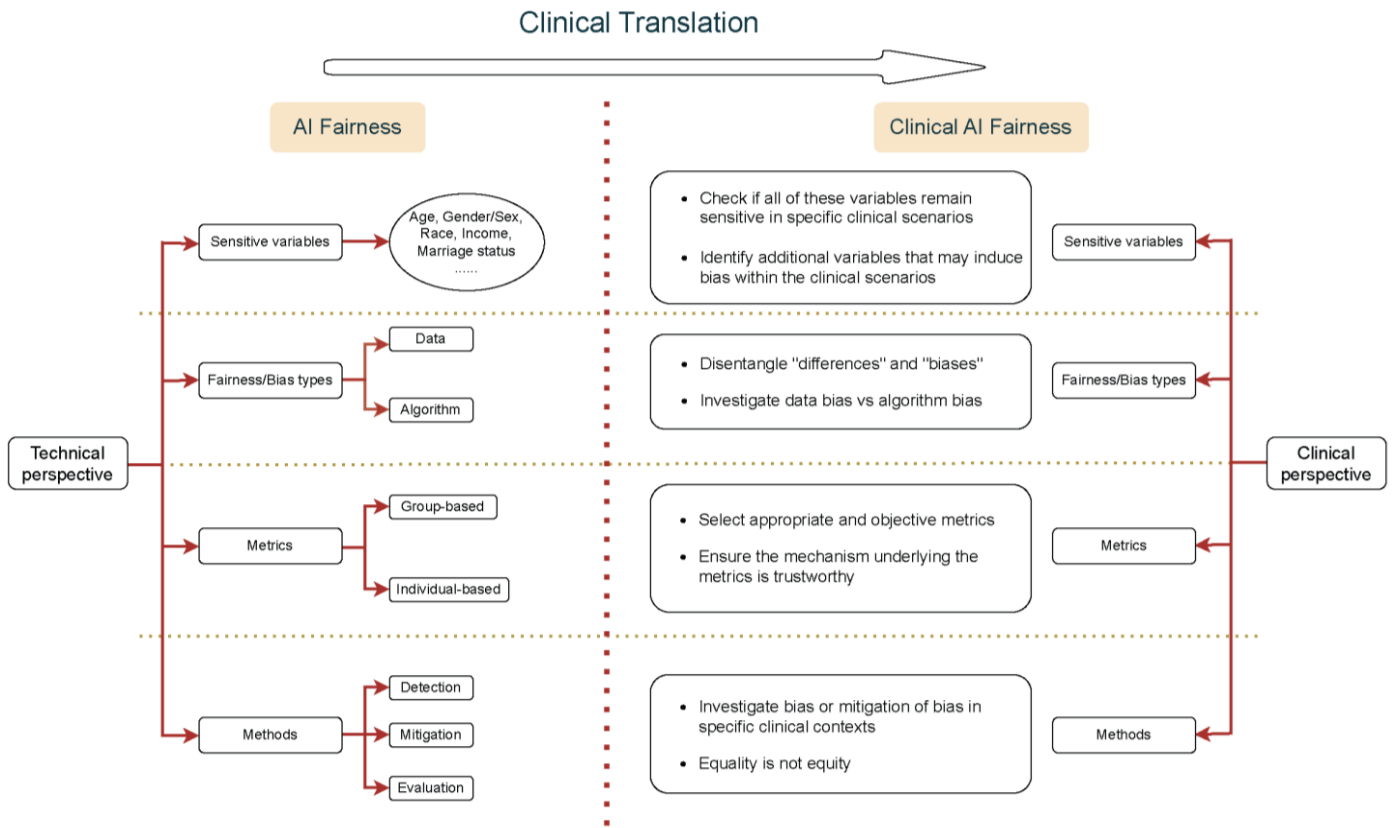